\documentclass[iop]{emulateapj}
\usepackage{amsmath}
\usepackage{float}
\usepackage{natbib}

\slugcomment{Accepted by the Astrophysical Journal}

\begin{document}

\title{AM Canum Venaticorum Progenitors with Helium Star Donors and the resultant Explosions}

\author{Jared Brooks\altaffilmark{1}, Lars Bildsten\altaffilmark{1,2}, Pablo Marchant\altaffilmark{3}, Bill Paxton\altaffilmark{2}}

\altaffiltext{1}{Department of Physics, University of California, Santa Barbara, CA 93106}
\altaffiltext{2}{Kavli Institute for Theoretical Physics, University of California, Santa Barbara, CA 93106}
\altaffiltext{3}{Argelander-Institut f{\"u}r Astronomie, Bonn, Germany, D-53121}

\begin{abstract}
We explore the outcome of mass transfer via Roche lobe overflow (RLOF) of $M_{\rm He}\lesssim0.51 M_\odot$ pure helium burning stars in close binaries with white dwarfs (WDs). 
The evolution is driven by the loss of angular momentum through gravitational wave radiation (GWR), and both stars are modeled using Modules for Experiments in Stellar Astrophysics (\texttt{MESA}). 
The donors have masses of $M_{\rm He}=0.35, 0.4, \&\ 0.51M_\odot$ and accrete onto WDs of mass $M_{\rm WD}$ from $0.6M_\odot$ to $1.26M_\odot$. 
The initial orbital periods ($P_{\rm{orb}}$) span 20 to 80 minutes.  
For all cases, the accretion rate onto the WD is below the stable helium burning range, leading to accumulation of helium followed by unstable ignition. 
The mass of the convective core in the donors is small enough so that the WD accretes enough helium-rich matter to undergo a thermonuclear runaway in the helium shell before any carbon-oxygen enriched matter is transferred.  
The mass of the accumulated helium shell depends on $M_{\rm WD}$ and the accretion rate.  
We show that for $M_{\rm He}\gtrsim0.4 M_\odot$ and $M_{\rm WD}\gtrsim0.8 M_\odot$, the first flash is likely vigorous enough to trigger a detonation in the helium layer.  
These thermonuclear runaways may be observed as either faint and fast .Ia SNe, or, if the carbon in the core is also detonated, Type Ia SNe. 
Those that survive the first flash and eject mass will have a temporary increase in orbital separation, but GWR drives the donor back into contact, resuming mass transfer and triggering several subsequent weaker flashes.
\end{abstract}


\section{Introduction}\label{sec:intro}

Core burning helium stars in semi-detached close binaries with WDs are one of the progenitor classes of AM Canum Venaticorum (AM CVn) stars, along with donors that are helium WDs and evolved main sequence stars.
They are part of a larger class of binary stars called cataclysmic variables, but are further defined by their short orbital periods (5-80 minutes) and a complete absence of hydrogen.
Their importance stems from common envelope stages during their formation, their detectability as gravitational wave sources, and their roles as possible projenitors for
helium novae, SNe .Ia, and even SNe Ia (\cite{1982ApJ...253..798N}, \cite{1982ApJ...257..780N}, \cite{Jr1991}, \cite{1994ApJ...423..371W}, \cite{2007ApJ...662L..95B}, \cite{2007MNRAS.381..525D}, \cite{Yungelson2008}, \cite{2008ApJ...684.1366K}, \cite{Shen2009}, \cite{2009MNRAS.395..847W}, \cite{2009ApJ...699.2026R}, \cite{2009MNRAS.400L..24S}, \cite{2011ApJ...734...38W}, \cite{2012ApJ...755....4T}, \cite{2014MNRAS.445.3239P}, \cite{2014MNRAS.440L.101R}, \cite{2014ApJ...785...61S}).

One route by which He star + WD binaries are formed is realized/encountered when the initial binary configuration leads to so-called case CB mass transfer: the primary fills its Roche lobe during its AGB phase (C-type), and after a common envelope phase, the secondary fills its Roche lobe during its RGB phase (B-type) (\cite{Jr1994}).
The two common envelope phases release a large fraction of the initial angular momentum, leaving a helium star in a short orbital period with a WD.
The binary is then driven closer together due to angular momentum loss via gravitational wave radiation until the helium star fills its Roche lobe and begins donating helium to the WD.
Longer initial orbital periods lead to helium stars that complete their burning before making contact.

When the WD has accumulated enough accreted helium, a thermonuclear runaway begins in the helium shell.
At this point, the WD will likely eject all the mass above the burning layer (\cite{Jr1991}).
Depending on the mass of the WD and of the accreted helium shell (and of the fraction of metals in the accreted material) the helium shell burning may result in a deflagration or detonation (\cite{1990ApJ...354L..53L}, \cite{1991ApJ...371..317L}, \cite{1991ApJ...370..272L}, \cite{1994ApJ...423..371W}, \cite{2007ApJ...662L..95B}, \cite{Shen2009}, \cite{2011ApJ...734...38W}, \cite{Moore2013}, \cite{Shen2014}, \cite{2014ApJ...785...61S}).
Helium detonations certainly unbind the shell and lead to fast and faint .Ia SNe.
They may also trigger a carbon detonation in the core, which causes a Type Ia SN (\cite{1990ApJ...354L..53L}, \cite{1991ApJ...370..272L}, \cite{1994ApJ...427..330A}, \cite{1995ApJ...452...62L}, \cite{2007A&A...476.1133F}, \cite{2010A&A...514A..53F}, \cite{2010ApJ...714L..52S}, \cite{2010ApJ...719.1067K}, \cite{2011ApJ...734...38W}, \cite{2012MNRAS.420.3003S}, \cite{2013ApJ...774..137M}, \cite{2014ApJ...785...61S}).

The goal of our work is to use a realistic mass transfer model to calculate the He shell masses at ignition and assess the likelihood of their detonations.
The mass loss from these events, which, given survival of the accretor, occur multiple times for each system, increases the orbital separation (at these low mass ratios), causing the stars to temporarily lose contact, until GWR drives them together again.
Therefore, while the initial combined mass of the two compact stars may exceed the Chandrasekhar limit, this channel will not result in a near Chandrasekhar mass WD.

In \S \ref{sec:donor} we describe the creation and subsequent simulation of the helium star models, compare the secular evolution results to those of previous studies, and explain the donor stars' response to mass loss during different stages.
Then in \S \ref{sec:wd} we include the behavior of the accreting WD and the conditions leading to the first thermonuclear runaway in order to determine the strength of the flash and the likelihood of a detonation occurring. 
As a case study, we simulated the binary system CD-30$^{\circ}$ 11223 (\cite{Geier2013}) up to the first flash.
Additionally, we follow through with the simulations assuming the first flash does not unbind the accreting WD, and calculate the effect of mass loss from ejection events on the binary parameters in \S \ref{sec:later}, and follow up with conclusions and discussion in \S \ref{sec:conc}.

\section{Donor Evolution and Mass Transfer Rates}\label{sec:donor}

These types of binary systems are formed when two comparable mass stars, with mass ratio less than 2, are born in a close binary. The more massive (primary) star will leave the MS and evolve through the RGB and AGB phases, donating some of its mass to its less massive companion. 
The primary star then becomes a WD, while the secondary, now the more massive of the two, accelerates through its MS evolution.
Once the secondary leaves the MS and expands as it rises up the RGB, it overflows its RL and begins unstable mass transfer.
The resulting common envelope removes the secondary's hydrogen envelope, revealing a low mass helium star.
This happens soon enough that the WD's core is still hot ($\log(T_c/\rm{K})>7.2$) at an age of {$\approx$}200 Myr.
The two stars are then close enough for GWR to drive them into contact before exhaustion of helium in the secondary's core as long as $P_{\rm{orb}}\lesssim 2$ hours after the final common envelope (\cite{1989SvA....33..606T}, \cite{1990SvA....34...57T}, \cite{Jr1991}, \cite{Jr1994}, \cite{Yungelson2008}). 

Our $0.35$ $M_\odot$ helium star model is created by evolving a $3.0$ $M_\odot$ ZAMS star with solar metallicity, using a mixing length of twice the pressure scale height ($\alpha_{ml} = 2H$) and the Schwarzschild criterion with no convective overshooting, through the main sequence and into hydrogen shell burning until the mass fraction of hydrogen at mass coordinate $M_r=0.35$ $M_\odot$ has dropped below $10^{-6}$.  
We then artificially remove mass from the surface  until the mass is $0.35$ $M_\odot$.
The star is left to adjust to the mass loss until core helium burning luminosity reaches 1 $L_\odot$.
The $0.4$ $M_\odot$ helium star is created the same way starting with a $3.5$ $M_\odot$ ZAMS star.
We also created a $0.51$ $M_\odot$ sdB star by evolving a $1$ $M_\odot$ ZAMS star up to the helium core flash, at which point we turn on a very strong RGB wind until the mass drops to $0.51$ $M_\odot$.
The core flash then proceeds to stable core He burning.

We set ``$\texttt{tau\_factor=100}$'', which puts the outer cell of the model at an optical depth of $\tau=(2/3)\times\texttt{tau\_factor}$.
This setting omits the HeII ionization zone and leads to a thin surface convection zone and gives a $T_{\rm eff}$ consistent with sdB star observations.
Setting a $\texttt{tau\_factor=1}$ leads to a deep surface convection zone, and initial mass transfer rates experience a runaway and encounter numerical problems.
We leave this issue for future studies.

\subsection{Mass Transfer Rates}\label{sec:mdot}

\begin{figure}[htb]
  \centering
  \includegraphics[width = \columnwidth]{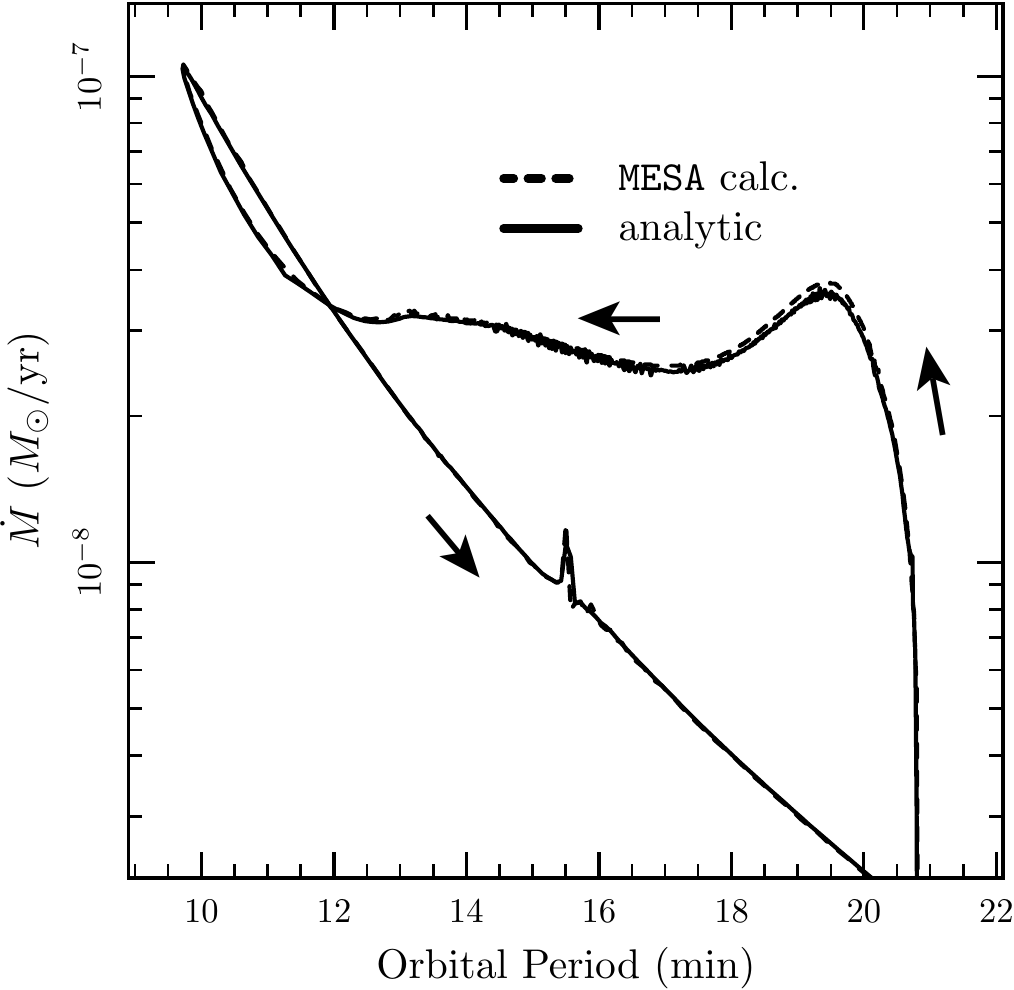}
  \caption{Comparison of $\dot{M}$ calculated by \texttt{MESA}'s root-finding algorithm (solid line) and using equations (\ref{eqn:3}) and (\ref{eqn:4}) given by component masses and $R_{\rm{He}}$ given in \texttt{MESA} output (dashed line). 
  Arrows show the direction of evolution.
  From the case with $M_{\rm He}=0.4 M_\odot$, $M_{\rm WD}=0.6 M_\odot$ at $P_{\rm{orb},0}=40$ minutes.}  
  \label{fig:1}
\end{figure}

\begin{deluxetable*}{ccccccccc}[H]
  \tablecaption{Comparison to previous work by Yungelson}
  \tablehead{\colhead{$M_{\rm He}/M_\odot$} & \colhead{$M_{\rm WD}/M_\odot$} & \colhead{$P_{\rm{orb},0}$ (min)} & \multicolumn{2}{c}{\underline{$t_c$ (Myr)}} & \multicolumn{2}{c}{\underline{$P_c$ (min)}}  & \multicolumn{2}{c}{\underline{$Y_c$}}\\ \colhead{} & \colhead{} & \colhead{} & \colhead{Yungelson} & \colhead{\texttt{MESA}} & \colhead{Yungelson} & \colhead{\texttt{MESA}} & \colhead{Yungelson} & \colhead{\texttt{MESA}}}\\
  \startdata
  0.35  & 0.5   & 20   & 1.29    & 1.4   & 15.96  & 15.7  & 0.977  & 0.974 \\
  0.35  & 0.5   & 40   & 15.99   & 17.4  & 16.24  & 16.1  & 0.936  & 0.923 \\
  0.35  & 0.5   & 60   & 50.80   & 51.1  & 17.02  & 16.9  & 0.871  & 0.833 \\
  0.4   & 0.8   & 40   & 9.14    & 9.2   & 20.61  & 21.2  & 0.933  & 0.925 \\
  0.4   & 0.8   & 60   & 30.21   & 30.4  & 21.67  & 22.0  & 0.854  & 0.824 \\
  0.4   & 0.8   & 80   & 66.95   & 67.4  & 22.85  & 24.1  & 0.751  & 0.637
  \enddata
  \label{tab:1}
  \tablecomments{$M_{\rm He}$ and $M_{\rm WD}$ are the initial masses of the donor and the accretor, $P_{\rm{orb},0}$ is the initial orbital period, $t_c$ is the time before the donor fills its Roche lobe, $P_c$ is the orbital period at time of contact, and $Y_c$ is the central Helium mass fraction at the time of contact.}
\end{deluxetable*}

The binary interactions are first computed, using \texttt{MESA} (\cite{2011ApJS..192....3P}, \cite{2013ApJS..208....4P}) version 5118, by adding a point-mass companion, i.e. the evolution of the companion star is not computed during this calculation.
We assume no rotation or magnetic braking, and conservative mass transfer.
The only factors that affect the binary separation, $a$, are mass transfer and loss of angular momentum via GWR.
To compute mass transfer rates, we used the ``Ritter'' implicit scheme of $\texttt{MESA}$ (Paxton et al. 2015 in preparation), which implicitly computes the prescription given by \cite{1988A&A...202...93R}.

The resulting mass transfer rate, $\dot{M}_{\rm He}$, matches what we would expect from the relation of mass and angular momentum loss rates,
\begin{equation}
\dfrac{\dot{M}_{\rm{He}}}{M_{\rm{He}}}\left(\dfrac{n}{2}+\dfrac{5}{6}-\dfrac{M_{\rm{He}}}{M_{\rm{WD}}}\right)=\dfrac{\dot{J}}{J}\bigg|_{GWR}\hspace{3mm},
\label{eqn:3}
\end{equation}

where $M_{\rm{He}}$ and $M_{\rm{WD}}$ are the masses of the helium star and WD, respectively, and $n=d\text{ln}R_{\rm{He}}/d\text{ln}M_{\rm{He}}$ (\cite{1967AcA....17..287P}, \cite{2001A&A...368..939N}).
Angular momentum loss via GWR is given by \cite{Landau1962} as
\begin{equation}
\dfrac{\dot{J}}{J}\bigg|_{GWR}=-\dfrac{32}{5}\dfrac{G^3}{c^5}\dfrac{M_{\rm{He}}M_{\rm{WD}}(M_{\rm{He}}+M_{\rm{WD}})}{a^4}\hspace{3mm}.
\label{eqn:4}
\end{equation}

\begin{figure}[H]
  \centering
  \includegraphics[width = \columnwidth]{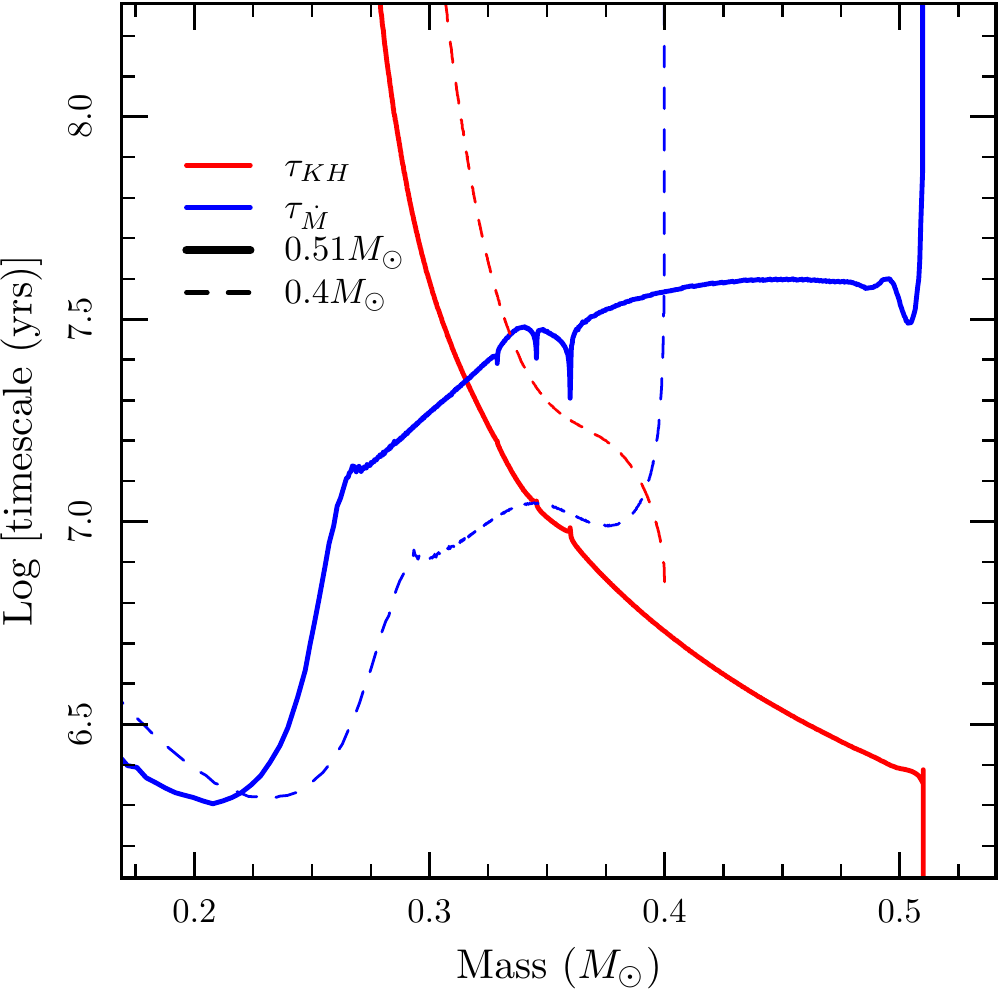}
  \caption{The two models shown here are the $0.51$ $M_\odot$ and the $0.4$ $M_\odot$ donors. 
  The two timescales for each model are the thermal and the mass loss timescales. 
  At the beginning of the mass transfer phase, the thermal timescale is much shorter than the mass loss timescale, so the donor star adjusts its thermal structure in response to mass loss. 
  As mass transfer rates rise and core luminosity falls, the timescales become comparable, and the donor responds adiabatically to mass loss, causing thermal timescales to rise and mass loss timescales to fall even faster.}
  \label{fig:6}
\end{figure}

\noindent Figure \ref{fig:1} shows the $\dot{M}$ computed by \texttt{MESA} and the $\dot{M}$ calculated by equations (\ref{eqn:3}) and (\ref{eqn:4}) against $P_{\rm{orb}}$ given component masses and $R_{\rm{He}}$ (for the derivative $n$) reported in \texttt{MESA} output for a case with $M_{\rm He}=0.4 M_\odot$, $M_{\rm WD}=0.6 M_\odot$ at $P_{\rm{orb},0}=40$ minutes.
The derivative $n=d\text{ln}R_{\rm{He}}/d\text{ln}M_{\rm{He}}$ was smoothed in certain regions using a raised cosine window because of numerical noise.
The donor first starts RLOF at $P_{\rm orb}\approx21$ minutes and $\dot{M}$ quickly rises to $\approx3\times10^{-8} M_\odot$/yr, follows the arrows to a period minimum at 10 minutes, and then expands to larger $P_{\rm orb}$ with a declining $\dot{M}$.
The glitch at ${\approx}15.5$ minutes results from the exposure of the once convective helium burning core.

In addition to checking $\dot{M}$ histories from \texttt{MESA} against equations (\ref{eqn:3}) and (\ref{eqn:4}), we also compare our results to the work of \cite{Yungelson2008}.
Shown in Table \ref{tab:1} are the age, $P_{\rm{orb}}$, and the core mass fraction of helium at the time of contact for six cases.
We agree with this prior work.
For a final comparison to Yungelson's results, we plotted the evolution of the thermal ($\tau_{\rm{KH}}=GM^2/RL$) and angular momentum loss ($\tau_{\rm{GW}}=(J/\dot{J})|_{GWR}$) timescales for the case with initial parameters $M_{\rm{He}}=0.35$ $M_\odot$, $M_{\rm{WD}}=0.5$ $M_\odot$, and $P_{\rm{orb},0}=20$ min and found good agreement with the same plot shown in Figure 3(a) in \cite{Yungelson2008}.

\subsection{Evolution of the Helium Star Donors}\label{sec:donor evol}

Figure \ref{fig:6} shows the evolution of the thermal and mass-loss ($\tau_{\dot{M}}=M/\dot{M}$, related to $\tau_{\rm GW}$ by equation \ref{eqn:3}) timescales for two cases with initial parameters ($M_{\rm{He}}$, $M_{\rm{WD}}$, $P_{\rm{orb},0}$) = ($0.51$ $M_\odot$, $0.76$ $M_\odot$, 70.5 min), ($0.4$ $M_\odot$, $1.0$ $M_\odot$, 40 min).
Models with lower donor mass climb to higher $\dot{M}$ faster because they make contact at smaller orbital separations, which means that radii and luminosities decrease faster and $\tau_{\rm{KH}}\approx\tau_{\rm{\dot{M}}}$ happens sooner in the evolution.
This timescale crossing happens almost immediately after mass transfer turn-on for the two smaller mass donors, while the model with initial mass $M_{\rm{He}}=0.51$ $M_\odot$ has $\approx0.2$ $M_\odot$ removed before $\tau_{\rm{KH}}\approx\tau_{\rm{\dot{M}}}$.

\begin{figure}[H]
  \centering
  \includegraphics[width = \columnwidth]{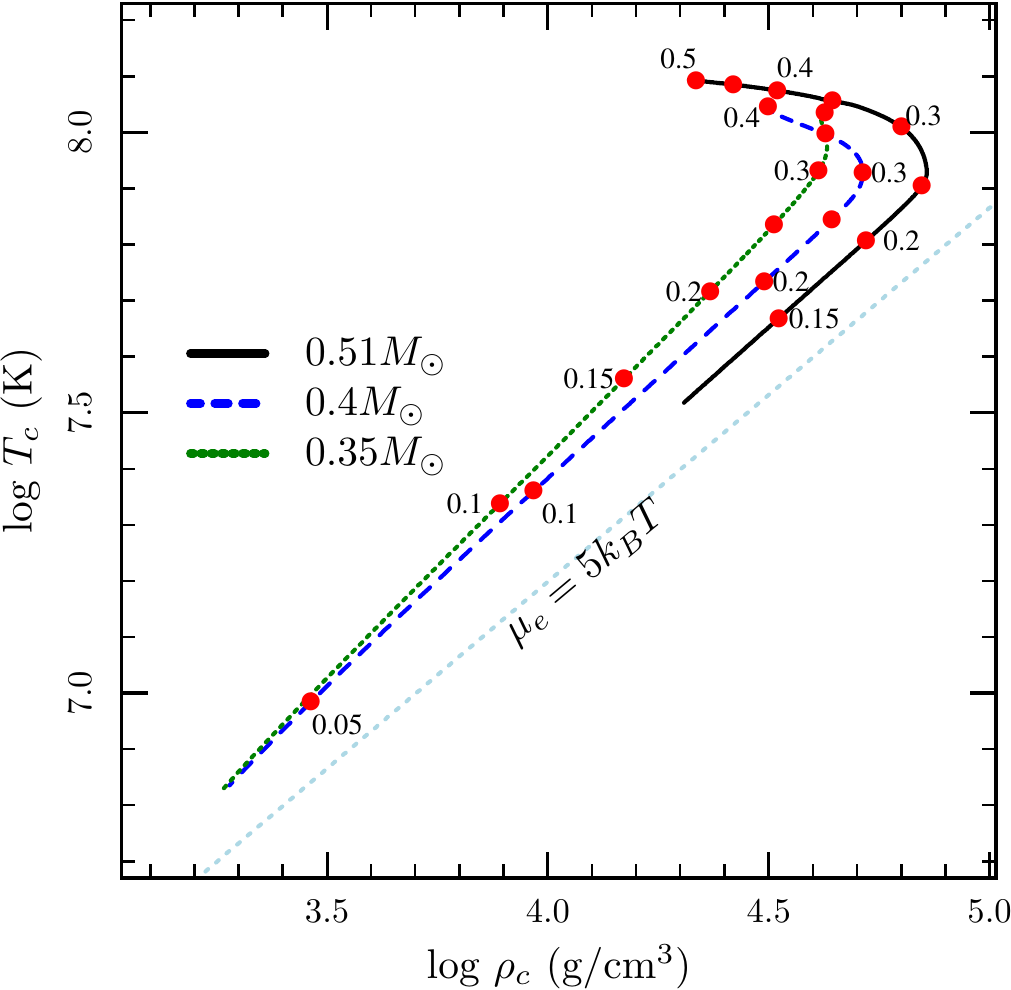}
  \caption{Evolutionary tracks of the central conditions in the donor stars. 
  The $M_{\rm{He}}=0.51$ $M_\odot$ case has a $M_{\rm{WD}}=0.76$ $M_\odot$ companion with $P_{\rm{orb},0}=70.5$ min and donor core electron degeneracy parameter $\eta_c=3.8$ (see end of sect. \ref{sec:donor evol}) when $M_{\rm{He}}=0.2$ $M_\odot$. 
  The $M_{\rm{He}}=0.4$ $M_\odot$ case has a $M_{\rm{WD}}=1.0$ $M_\odot$ companion with $P_{\rm{orb},0}=40$ min and donor core electron degeneracy parameter $\eta_c=3.1$ when $M_{\rm{He}}=0.2$ $M_\odot$. 
  The $M_{\rm{He}}=0.35$ $M_\odot$ case has a $M_{\rm{WD}}=1.0$ $M_\odot$ companion with $P_{\rm{orb},0}=20$ min and donor core electron degeneracy parameter $\eta_c=2.6$ when $M_{\rm{He}}=0.2$ $M_\odot$.}
  \label{fig:2}
\end{figure}

\begin{figure}[H]
  \centering
  \includegraphics[width = \columnwidth]{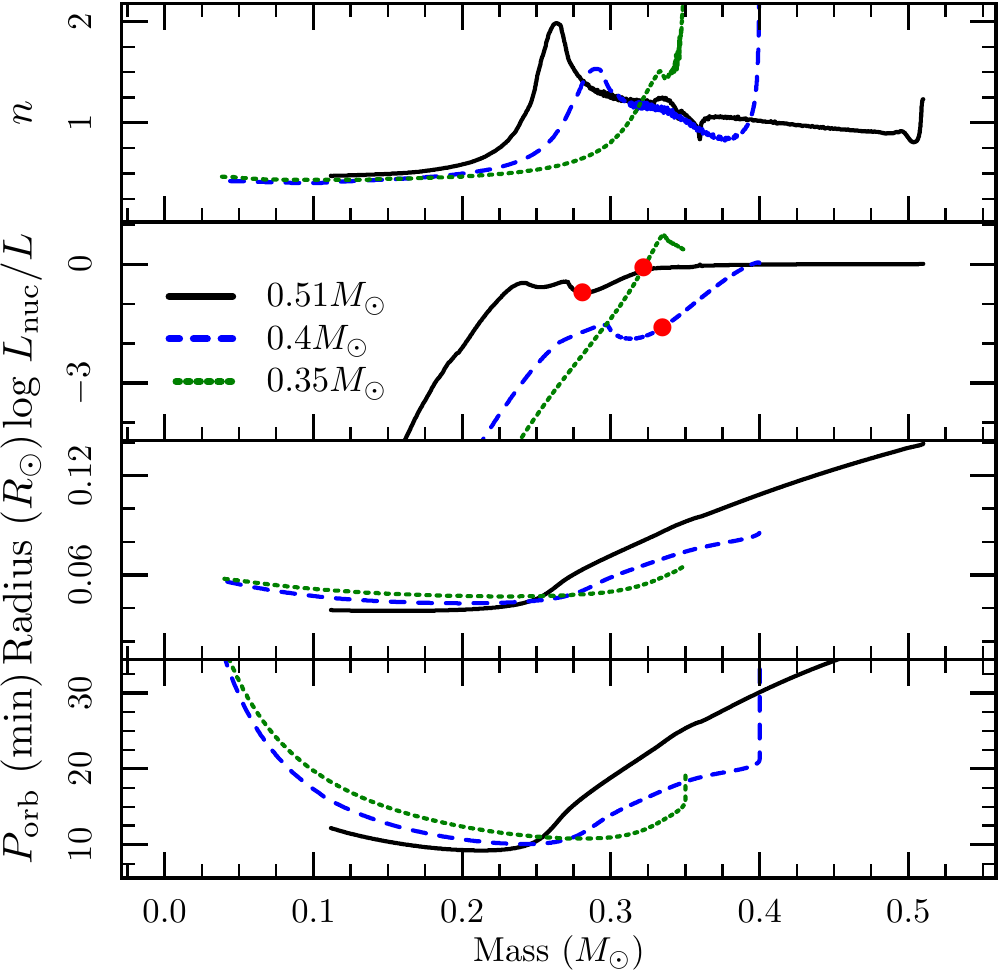}
  \caption{The models here are the same as plotted in Figures \ref{fig:2} and \ref{fig:6}. 
  The top section shows the derivative $n=d\text{ln}R_{\rm{He}}/d\text{ln}M_{\rm{He}}$, which was smoothed. 
  The red circles in the middle panel mark when the core convection zone disappears.
  The mass at which the period minimum is reached for each model are ($M_{\rm{initial}}$, $M_{P_{\rm{orb,min}}}$) = ($0.51$ $M_\odot$, $0.21$ $M_\odot$), ($0.4$ $M_\odot$, $0.24$ $M_\odot$), and ($0.3$ $M_\odot$, $0.28$ $M_\odot$), which corresponds to $n\approx0.14$ for each case.}
  \label{fig:5}
\end{figure}

As the nuclear burning luminosity starts to decrease due to mass loss, the core temperature slowly decreases, allowing core contraction to increase the central density, shown in Figure \ref{fig:2}.
The bump in $L_{\rm{nuc}}/L$ in the middle panel of Figure \ref{fig:5} is caused by the envelope's absorption of the nuclear burning luminosity due to mass-loss (\cite{Savonije1986},\cite{Yungelson2008}), explained more in \S \ref{sec:later}.
After the bump in the top and middle panel of Figure \ref{fig:5}, $n$ begins to fall, which, as can be seen from equation (\ref{eqn:3}), increases $\dot{M}$. 
At this point the star is being driven out of thermal equilibrium as $\tau_{\rm{KH}}\gg\tau_{\rm{\dot{M}}}$, so further mass loss causes the central density to decrease, hence the bend in the $\rho_c-T_c$ curves in Figure \ref{fig:2} (\cite{Savonije1986}, \cite{2007MNRAS.381..525D}).

The increase in $\dot{M}$ continues until the period minimum is reached, where $n=d\text{ln}R_{\rm{He}}/d\text{ln}M_{\rm{He}}\approx0.14$ for each model.
The decreasing central temperature and density keep the star evolving along lines of nearly constant entropy where the cores are only mildly degenerate (\cite{Savonije1986}, \cite{2007MNRAS.381..525D}).
The core electron degeneracy parameter, $\eta_c=\mu_{e,c}/k_B T_c$, that each of the three models considered here reach when their masses drop to $0.2$ $M_\odot$ are ($M_{\rm{He}}$, $\eta_c$) = ($0.51$ $M_\odot$, $3.8$), ($0.4$ $M_\odot$, $3.1$), ($0.35$ $M_\odot$, $2.6$).
The fact that the donors are only mildly degenerate means that they will have higher mass transfer rates for a given $P_{\rm orb}$ than their fully degenerate counterparts during the AM CVn stage (\cite{2007MNRAS.381..525D}).

\section{Accretion onto White Dwarfs and the First Flash}\label{sec:wd}

As mentioned above, enough helium is accreted onto the WD for a thermonuclear runaway to occur.
A negligible amount of helium is burned before the runaway occurs, as the stable helium burning range for WDs of this mass is $10^{-6}M_\odot$/yr$\lesssim\dot{M}\lesssim 3\times10^{-6}M_\odot$/yr (\cite{2014MNRAS.445.3239P}, Brooks et al. 2015 in preparation), a full order of magnitude larger than the accretion rates achieved by the cases in this study.
The outcome may either be dynamical or hydrostatic, with the WD slowly expanding to fill its Roche lobe.  
Under the assumption that ejecta take the specific angular momentum of the accretor, mass loss causes the binary separation to increase, which temporarily shuts off mass transfer until GWR brings the component stars back into contact.
We chose not to explore the possibility that dynamical friction within expanding envelope removes extra angular momentum (\cite{2015arXiv150205052S}).
Below we show the results of ``true'' binary runs where both stars are evolved simultaneously, including the binary parameters, using \texttt{MESA} version 6596.
The WD models were created 

\begin{figure}[H]
  \centering
  \includegraphics[width = \columnwidth]{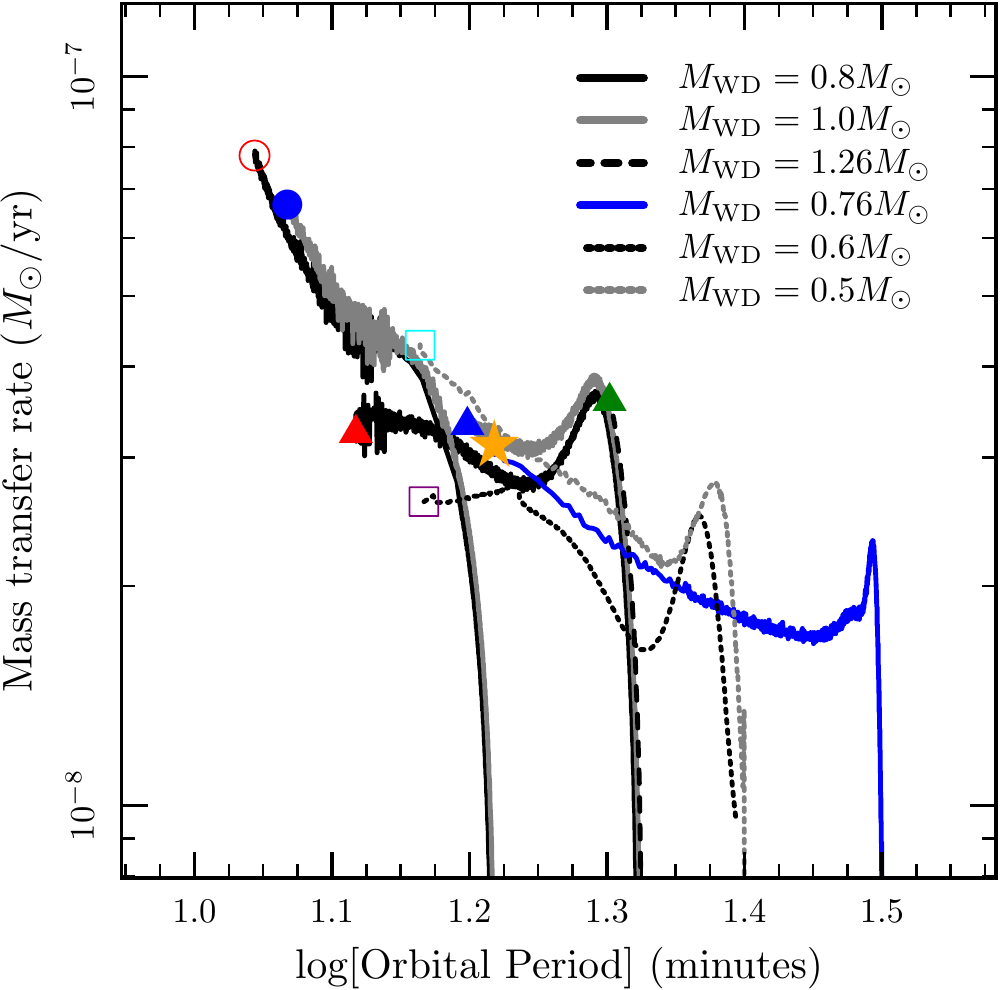}
  \caption{Black solid lines are systems with $M_{\rm WD}=0.8 M_\odot$, Grey solid lines are systems with $M_{\rm WD}=1.0 M_\odot$.
  The black dashed line is from a system with $M_{\rm WD}=1.26 M_\odot$.
  The black dotted line is from a system with $M_{\rm WD}=0.6 M_\odot$.
  The grey dotted line is from a system with $M_{\rm WD}=0.5 M_\odot$.
  The blue line is from the system modelling CD-30$^{\circ}$ 11223, see \S \ref{sec:case}.
  The symbol marking the first flash correspond with those in Figure \ref{fig:8}}
  \label{fig:14}
\end{figure}

\noindent in \texttt{MESA} by the same method described in \cite{Wolf2013}.
We age the WD models to 200 Myr, the relevant age for the 0.35 and 0.4 $M_\odot$ helium star donors.
These WDs have core temperatures of a few$\times10^7$K at the start of mass transfer.
The WDs with sdB star companions could be even colder.

In this section we show results from cases with ($M_{\rm He}$, $M_{\rm WD}$) = ($0.35 M_\odot$, $0.8 M_\odot$), ($0.35 M_\odot$, $1.0 M_\odot$), ($0.4 M_\odot$, $0.8 M_\odot$), ($0.4 M_\odot$, $1.0 M_\odot$), ($0.4 M_\odot$, $1.26 M_\odot$), ($0.46 M_\odot$, $0.5 M_\odot$), ($0.46 M_\odot$, $0.6 M_\odot$) all with initial orbital periods of $P_{\rm orb,0} = 40$ minutes, and a case with ($0.51 M_\odot$, $0.76 M_\odot$) and $P_{\rm orb,0} = 70.5$ to model the observed system CD-30$^{\circ}$ 11223.

\begin{figure}[H]
  \centering
  \includegraphics[width = \columnwidth]{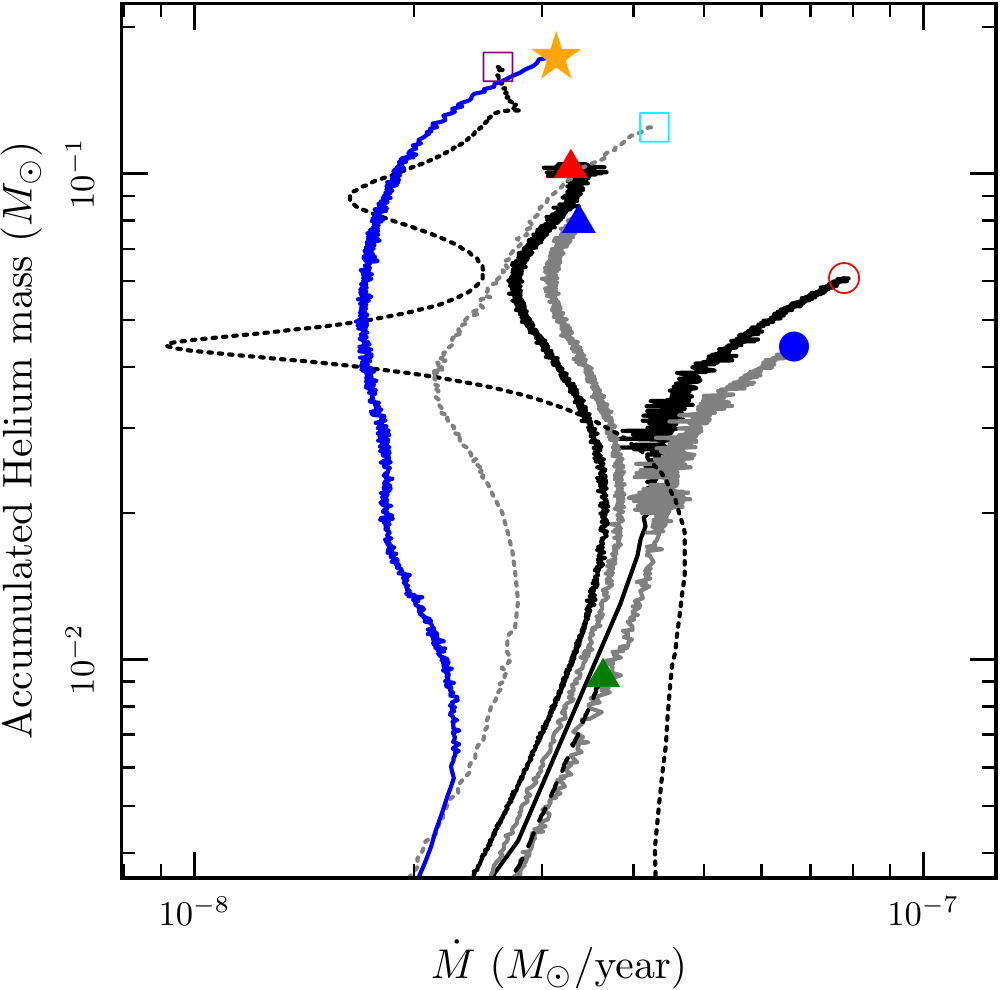}
  \caption{Accumulated helium shell mass for the first flash. 
  Line Types, colors, and markers are the same as for Figure \ref{fig:14}.}
  \label{fig:15}
\end{figure}

\begin{figure}[H]
  \centering
  \includegraphics[width = \columnwidth]{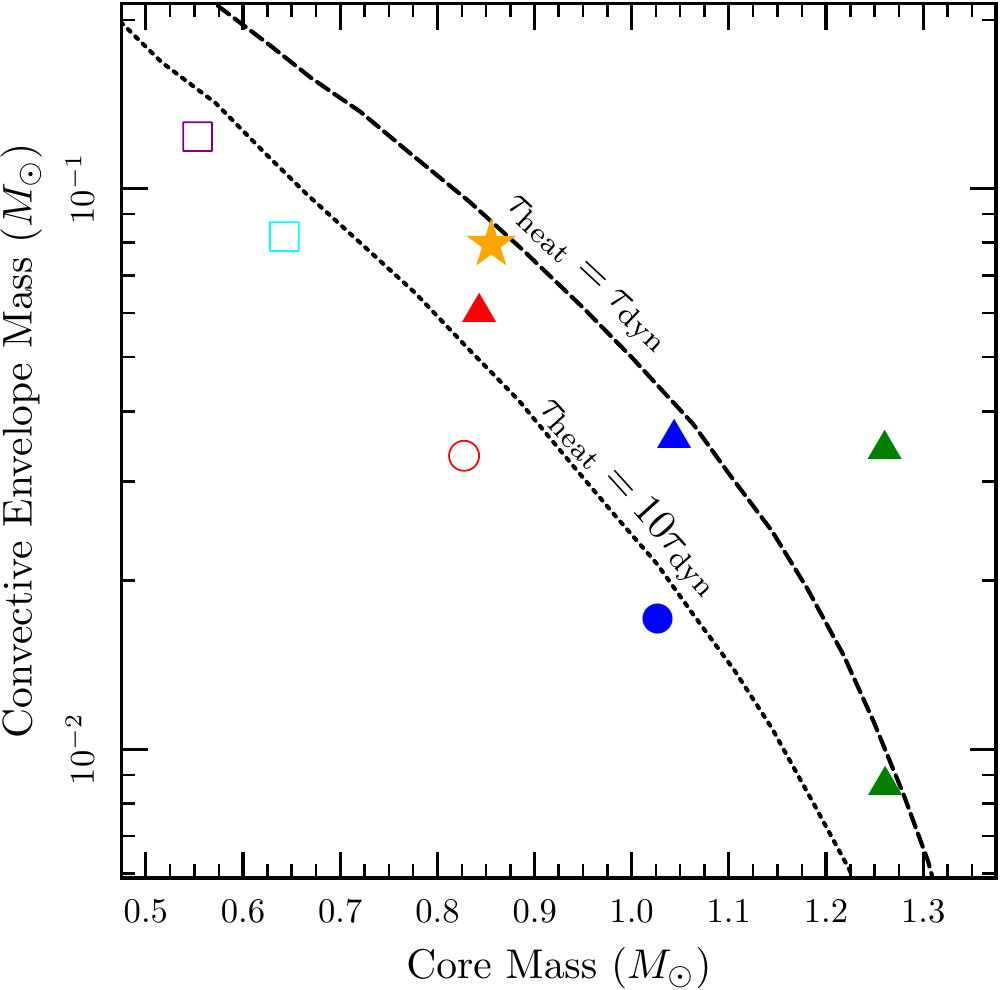}
  \caption{First burst from all ``true'' binary runs. 
  Blue markers are from $1.0 M_\odot$ WDs, red markers from $0.8 M_\odot$ WDs, green from $1.26 M_\odot$ WDs, purple from $0.5 M_\odot$ WDs, and cyan from $0.6 M_\odot$ WDs. 
  Circles are from $0.35 M_\odot$ donors, triangles from $0.4 M_\odot$ donors, squares from $0.46 M_\odot$ sdB star donors. 
  Filled markers had $c_pT/\epsilon_{\rm nuc}<\pi r/C_{\rm sound}$. 
  The orange star-shape is from the case study CD-30$^{\circ}$ 11223.
  The last (larger) burst on the $1.26 M_\odot$ WD is included as a special case and is discussed in \S \ref{sec:later}.}
  \label{fig:8}
\end{figure}

Our immediate goal is to calculate the thermal properties of the accumulating helium shell on the WD and find the shell mass at the time of the thermonuclear runaway.  
Most earlier work (see Figure 1 of \cite{Shen2009}) assumed a constant accretion rate, whereas these binaries undergo a factor of about three change in $\dot M$ during the accumulation phase. 
Figure \ref{fig:14} shows the changing $\dot M$ as the orbital period shrinks for our distinct cases. 
We terminate this plot at the moment of the first helium shell thermonuclear runaway. 
The resulting helium shell masses at the time of ignition (shown in Figure \ref{fig:15})  are not that different than earlier works, but are now derived self-consistently. 
At these accretion rates, there is adequate time for thermal contact with the colder core, resulting in a temperature inversion in the helium shell and an ignition location above the base of the freshly accreted material. 
Hence, the convective shell masses will always be less than the accumulated masses and we do not concern ourselves with possible mixing during the burning into the underlying CO (or ONeMg) core. 
\cite{2015ApJ...801..137P} recently discussed mixing during accumulation, which also looks unlikely at these lower accretion rates. 
Hence, we assume our shells have the abundance distribution of the material from the surface of the donor.

To calculate the nuclear burning rates we used \texttt{MESA}'s \texttt{co\_burn} network, which includes alpha chain reactions up to $^{28}$Si.
The qualitative nature of all bursts are quite similar, so we study one typical case in detail in \S \ref{sec:ex}. 

In Figure \ref{fig:8} we plot the core mass, which includes all the helium below the burning zone, against the convective envelope mass.
The black lines, from \cite{Shen2009}, represent where the local heating timescale ($c_pT/\epsilon_{\rm nuc}$) equals the (dashed line) local dynamical time ($H/C_{\rm sound}$), and (dotted) $10\times$ the local dynamical time, included for uncertainty in detonation requirements.
The filled markers met the condition that the burning timescale became shorter than the sound travel time across a hemisphere ($c_pT/\epsilon_{\rm nuc}<\pi r/C_{\rm sound}$), meaning that the burning was at least non-spherically symmetric.

\subsection{Typical First Helium Thermonuclear Instability}\label{sec:ex}

\begin{figure}[H]
  \centering
  \includegraphics[width = \columnwidth]{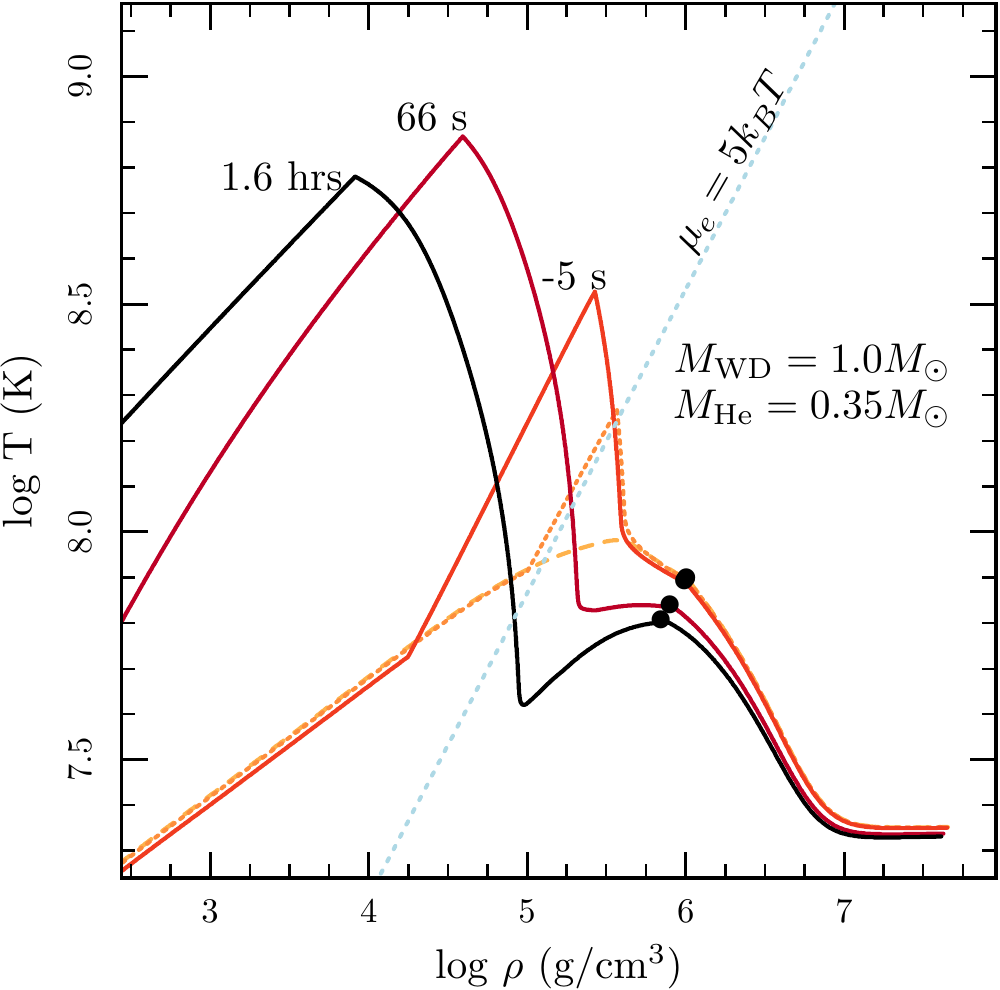}
  \caption{Temperature-density profiles from the specific helium explosion studied in this section. 
  The $0.0436 M_\odot$ of helium sits on a $1.0 M_\odot$ C/O core. 
  The size of the convective envelope (all the mass above the burning zone) is $0.0171 M_\odot$. 
  The first profile is dashed (1530 yrs before min. $\tau_{\rm heat}$), the second profile is dotted (0.5 yrs before min. $\tau_{\rm heat}$), and the third, fourth and fifth are labelled above by time since min. $\tau_{\rm heat}$.
  The black markers show the base of the helium shell in each profile (helium mass fraction $Y<0.05$).
  The light blue dotted line marks the degeneracy border where $\mu_e=k_BT$.}
  \label{fig:11}
\end{figure}

The explosion studied here is the first from the case with $M_{\rm He}=0.35 M_\odot$, $M_{\rm WD}=1.0 M_\odot$ (filled blue circle in Figure \ref{fig:8}).
Because the compressional heat leaks both out of the surface and into the core, the maximum temperature and burning zones are located within the helium shells, instead of the core-envelope boundary.
This ``off-center'' ignition was also found to be prevalent at constant $\dot{M}$ (\cite{1982ApJ...253..798N}, \cite{Jr1991}, \cite{2011ApJ...734...38W}).
The base of the helium shell (where helium mass fraction $Y<0.05$), shown by the black markers in Figure \ref{fig:11}, therefore, lies below the ignition location, and stays fairly degenerate.

\begin{figure}[H]
  \centering
  \includegraphics[width = \columnwidth]{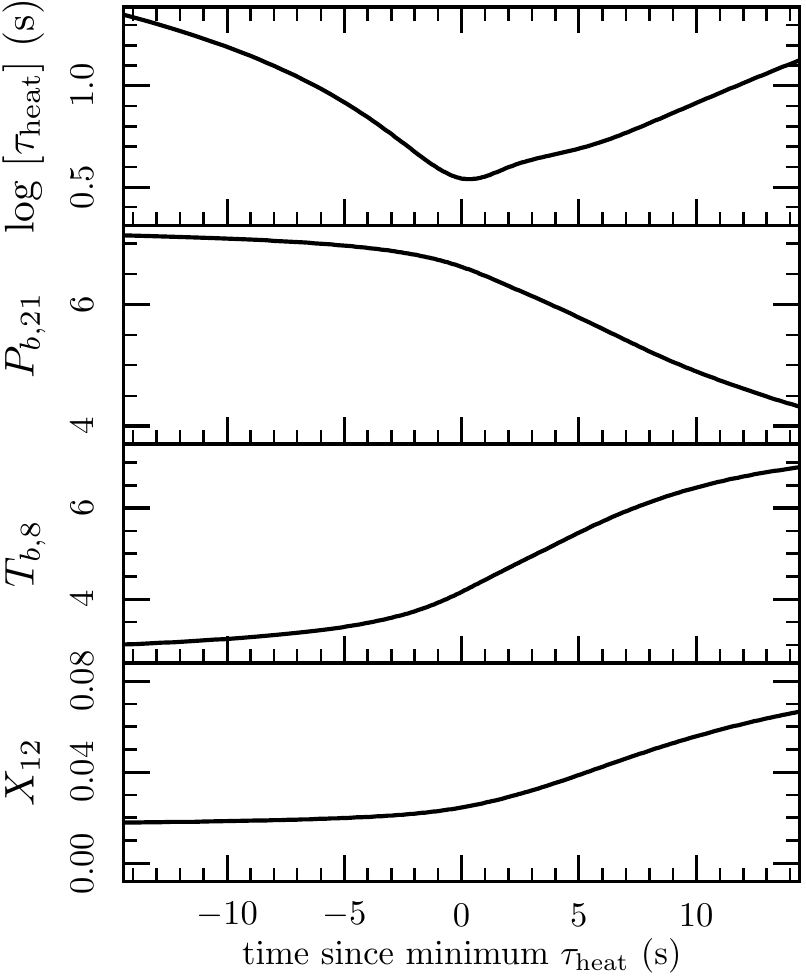}
  \caption{Same model as Figure \ref{fig:11}. 
  Shown are the evolution of four quantities defined at the maximum burning zone, with $t=0$ corresponding to the minimum $\tau_{\rm heat}$.
  The top row shows the local heating timescale, $\tau_{\rm heat}=c_pT/\epsilon_{nuc}$.
  The second row shows the local pressure, $P_{b,21}=P_b/10^{21}$erg$\cdot$ cm$^{-3}$.
  The third row shows the local temperature, $T_{b,8}=T_b/10^8$K.
  The bottom row shows the local mass fraction of $^{12}$C, $X_{12}$ }
  \label{fig:12}
\end{figure}

The evolution of the instability proceeds most rapidly near the minimum local heating timescale ($\tau_{\rm heat}=c_pT/\epsilon_{\rm nuc}$).
Shown in Figure \ref{fig:12}, at the base of the convective envelope (coincident with the burning zone) the temperature increases and the pressure decreases by a factor of ${\approx}2$, and the carbon mass fraction ($X_{12}$) rises from 0.02 to 0.08.
The heating timescale, $\tau_{\rm heat}$, reaches a minimum of ${\approx}3.5$ seconds, compared to a dynamical time ($\tau_{\rm dyn}=H/C_{\rm sound}$) of ${\approx}0.2$ seconds, and the sound travel time around a hemisphere, $\pi r/C_{\rm sound}\approx6$ seconds.
Convection extends almost all the way to the surface, and becomes fairly inefficient with convective speeds reaching ${\approx}15\%$ of the sound speed and eddy-turnover times ($\tau_{\rm conv}=H/v_{\rm conv}$) drop below 1 second, faster than the heating timescale, $\tau_{\rm heat}$.
This particular flash does not become dynamical, as dynamical flash models could not be evolved passed a $\tau_{\rm heat}$ minimum.

\begin{figure}[H]
  \centering
  \includegraphics[width = \columnwidth]{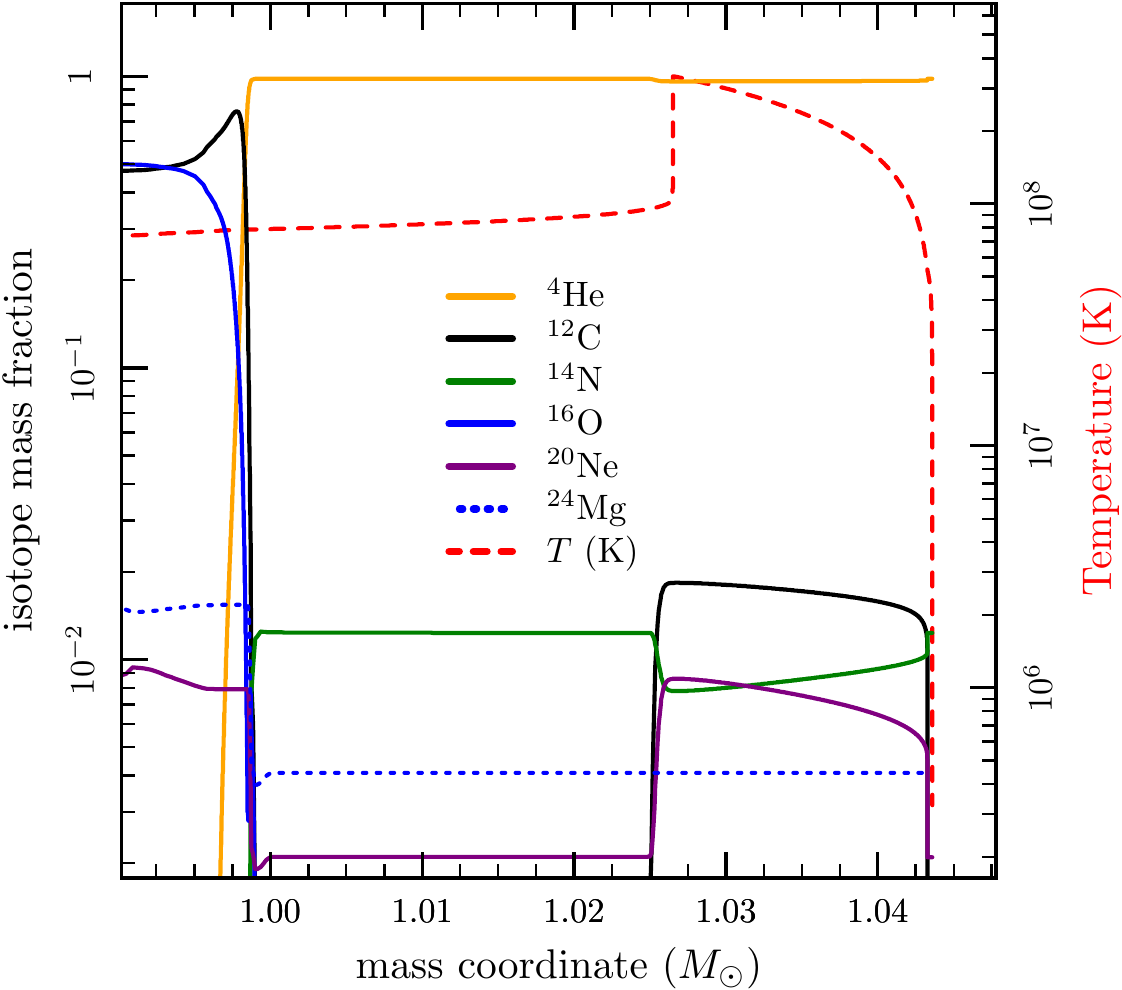}
  \caption{Composition and temperature profile of the helium envelope for the first burst of the system with $M_{\rm He}=0.35 M_\odot$, $M_{\rm WD}=1.0 M_\odot$ near minimum $\tau_{\rm heat}$}
  \label{fig:10}
\end{figure}

In Figure \ref{fig:10} we plot a composition profile of the helium envelope.
The accreted material is almost pure helium with ${\approx}1\%$ $^{14}$N, and the pre-dynamical convective burning produces ${\approx}2\%$ $^{12}$C and negligible $^{16}$O.

As the results of Figures \ref{fig:8} and \ref{fig:12} show, these helium shells evolve rapidly and we certainly should not depend on a hydrostatic code to resolve any transition to a more dynamical outcome such as a deflagration or detonation. 
However, we have shown that many of these first flashes (especially those with $M_{\rm He}>0.4M_\odot$ and $M_{\rm WD}>0.8M_\odot$) have an adequately massive convecting helium shell to trigger dynamical burning (i.e. deflagration or detonation).
In light of more recent work, specifically \cite{Shen2014}, who showed that using a larger nuclear reaction network that allows the proton-catalyzed $\alpha$-capture $^{12}$C$(p,\gamma)^{13}$N$(\alpha,p)^{16}$O significantly reduces that effective $^4$He lifetime, detonations are able to begin and propagate in much lower mass shells.
This and other reaction rates depend on the relative abundances in the burning layer.
Therefore we need to know not only the composition of the accreted matter, but also the build-up of burning products from pre-dynamical convective burning.

With the composition information from Figure \ref{fig:10}, which, as we stated above, is typical of the bursts studied here, we can plot each of these bursts on Figure 10 of \cite{Shen2014}, which tells that all the bursts studied here will be able to laterally propagate helium detonations.
Furthermore, it tells us that the largest bursts from the $0.8 M_\odot$ and the $1.0 M_\odot$ WD accretors, and all the burst from the $1.26 M_\odot$ WD accretor, will produce a significant fraction of isotopes that are radioactive on relevant timescales ($X_{\rm48Cr}+X_{\rm52Fe}+X_{\rm56Ni}>0.2$).
The rest of the smaller bursts will be rich in $^{28}$Si and $^{40}$Ca, but will mainly consist of unburnt helium, which will likely be neutral and unobservable.

\subsection{Case Study: CD-30$^{\circ}$11223}\label{sec:case}

The system CD-30$^{\circ}$ 11223 is the only known sdB+WD binary system expected to make contact within the sdB's core He-burning lifetime.
We used data measured by \cite{Geier2013} to model this system, which includes a $0.51 M_\odot$ sdB star and a $0.76 M_\odot$ WD in an orbital period of 70.5 minutes.
Comparison of measurements to simulated sdB evolutionary tracks in the $T_{\rm eff}-\log g$ diagram suggest that the sdB had just recently been formed and started the core He-burning phase.

Given the longer initial orbital period of this system compared to the systems in this study, the sdB had a longer time for core He-burning to inject heat into the envelope, leaving the star in a higher entropy state when mass transfer starts.
This leads to a lower mass transfer rate during the plateau phase by a factor of a few.
The lower mass transfer rate allows a larger helium shell to build up on the WD before thermonuclear runaway occurs.

\begin{figure}[H]
  \centering
  \includegraphics[width = \columnwidth]{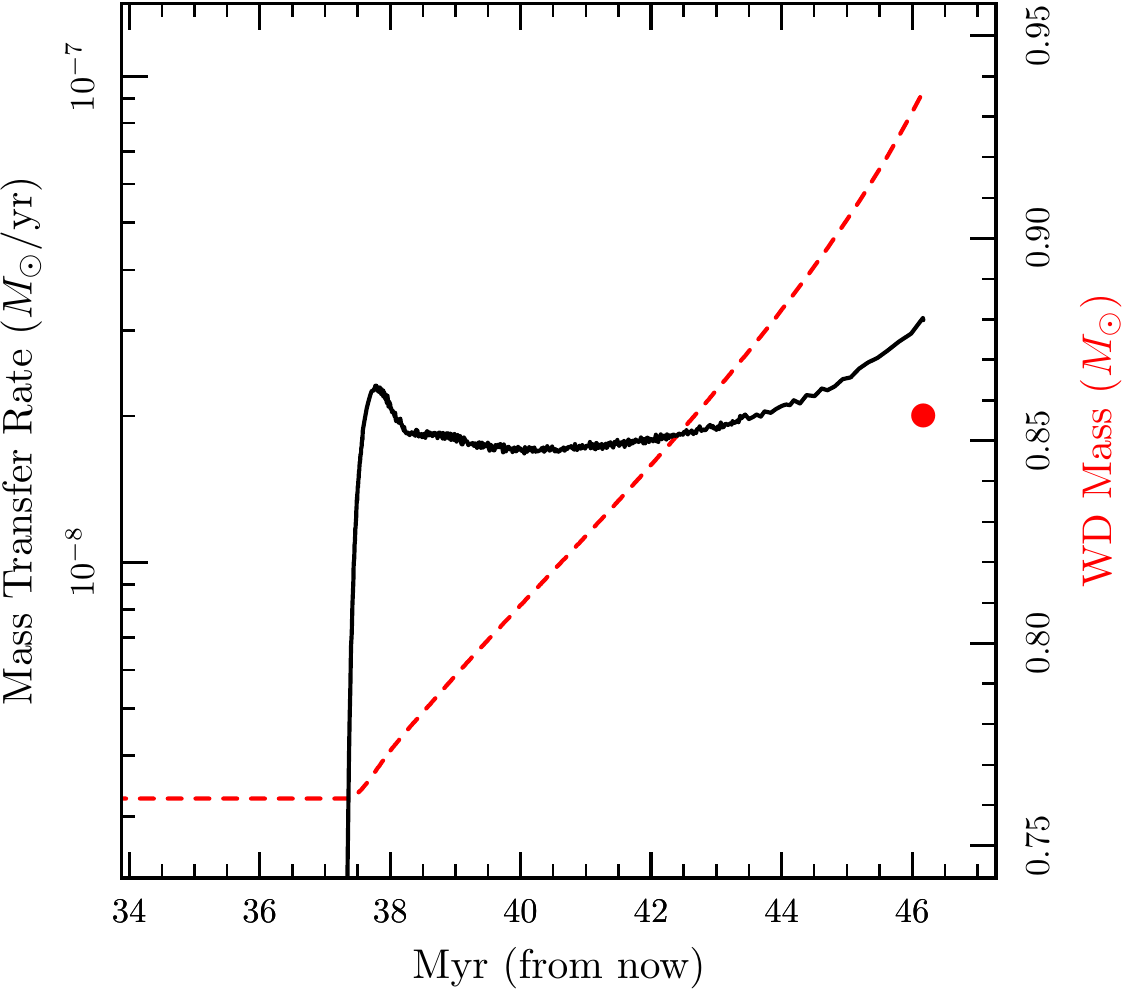}
  \caption{Mass transfer rate shown in black solid line, left y-axis.
  Mass of the WD accretor in red dashed line and mass coordinate of the ignition location as a red marker, right y-axis.}
  \label{fig:13}
\end{figure}

Our results from modeling this system show (Figure \ref{fig:13}) that the helium star fills its Roche lobe after 37.3 Myr at an orbital period of 32 minutes.
The donor transfers $0.18 M_\odot$ of helium to the WD before thermonuclear runaway after 46.1 Myr at an orbital period of 16.5 minutes.
Compare this to the prediction of \cite{Geier2013} of a helium shell of $0.1 M_\odot$ that explodes after only 42 Myr at an orbital period of 27.4 minutes.
The initial mass transfer rates we computed agree well with those in \cite{Geier2013}, but our model climbs to higher accretion rates by about 50\% by the first burst.

Our model shows a $0.09 M_\odot$ convective envelope on top of a ($0.76 M_\odot$ C-O + $0.09 M_\odot$ cold He) $0.85 M_\odot$ core, which will detonate and produce a significant mass of radioactive isotopes.

\section{Later Flashes and Subsequent Evolution}\label{sec:later}

If we assume that the system survives the first flash, the loss of mass via the ejection of the helium envelope will cause the binary separation to increase. 
GWR will then bring the component stars back into contact, and the cycle of helium accumulation, ignition, and ejection continues until the mass transfer rates fall to such low values that another ignition never occurs (\cite{Shen2009}).
To model this, each time helium burning on the WD begins to run away, the helium envelope is removed from the system, taking with it the specific orbital angular momentum of the WD.

\begin{figure}[H]
  \centering
  \includegraphics[width = \columnwidth]{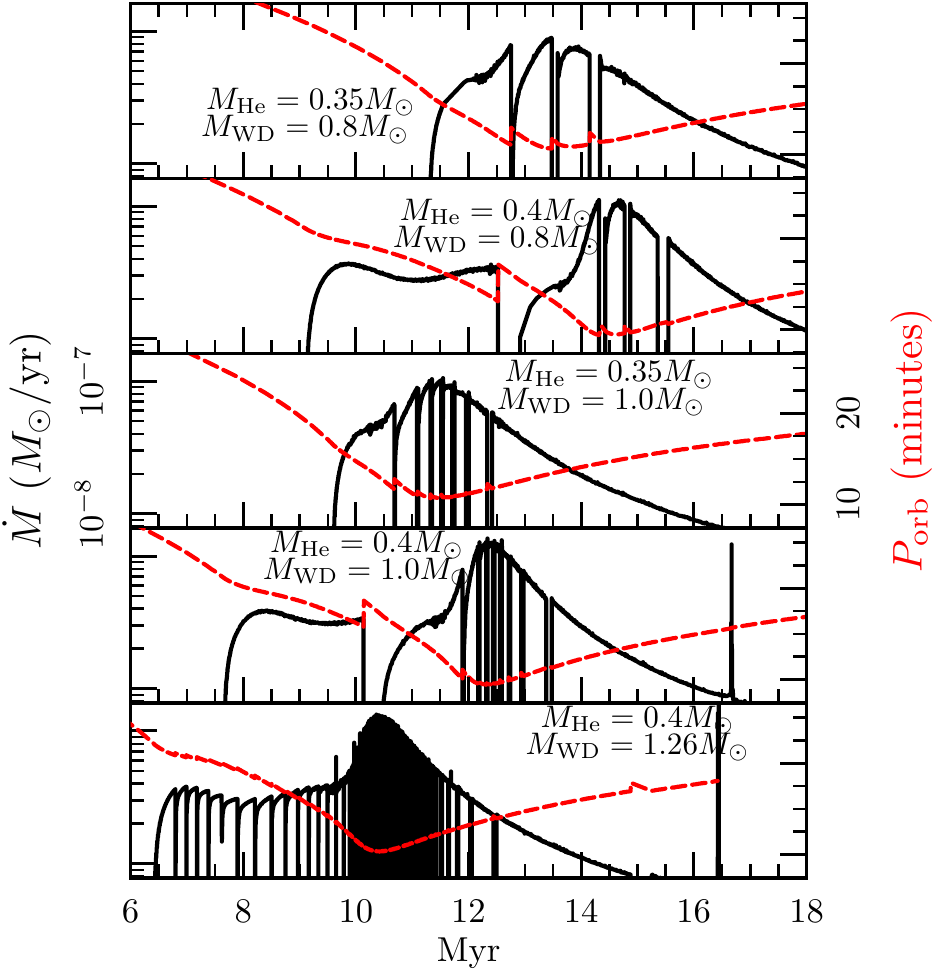}
  \caption{Four true binary runs all started at $P_{\rm{orb},0}=40$ minutes. The WD accretors experience between 3 and 9 bursts, each of which temporarily shut-off mass transfer. Mass transfer rates are shown by the black solid lines and the orbital periods are shown by the red dashed lines. All panels have the same scale.}
  \label{fig:3}
\end{figure}

\begin{figure}[H]
  \centering
  \includegraphics[width = \columnwidth]{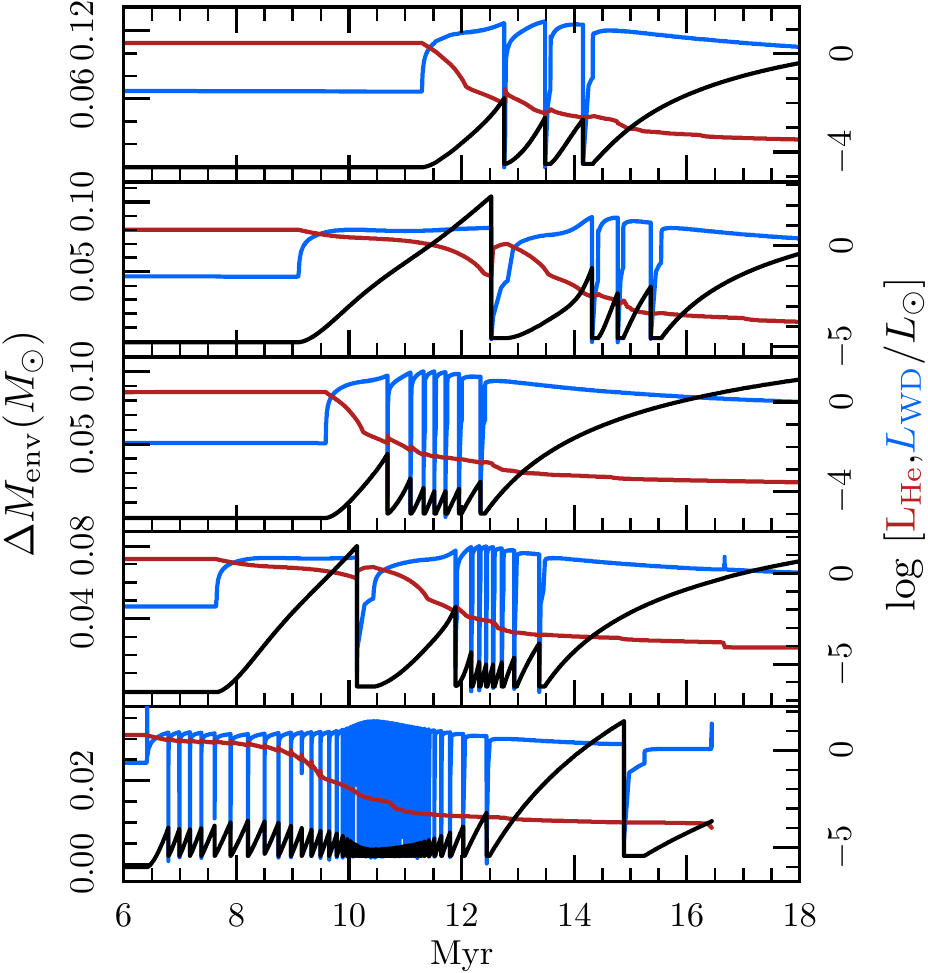}
  \caption{Same four binary runs as in Figure \ref{fig:3}. Mass of the helium envelope on the WD ($\Delta M_{\rm env}$) shown in black lines, luminosities of the helium star and WD in red and blue, respectively.}
  \label{fig:7}
\end{figure}

We show the effects on the mass transfer and orbital period histories caused by including the mass ejection episodes in Figure \ref{fig:3}.
The $0.35 M_\odot$ donor models lack the plateau feature in the $0.4 M_\odot$ donor models because the thermal timescale ($\tau_{\rm{KH}}$) drops below the mass transfer timescale ($\tau_{\dot{M}}$) much sooner for lower mass donors.
The higher mass WDs have lower envelope masses ($\Delta M_{\rm{env}}$) necessary for helium ignition, so these models have more bursts.
This behavior can also be seen in Figure \ref{fig:7}, which shows the mass of helium shell in the black lines, and the luminosities of the helium star and WD in red and blue, respectively.

Before contact, the helium star is more luminous than the WD.
After contact, that relation is reversed, and $L_{\rm{WD}}\sim10 L_\odot$.
This approximately matches the compression luminosity, $L_{\rm{comp}}\approx3k_BT_c\langle\dot{M}\rangle/\mu m_p$, given in \cite{2006ApJ...640..466B}.
This does not include the accretion luminosity, $L_{\rm{acc}}\approx GM_{\rm WD}\dot{M}/R_{\rm WD}\sim100 L_\odot$, which is released in the accretion disk and will outshine both stars.

For bursts that remove enough mass and mass transfer is shut off for long enough, the WD can be seen to dim to the original cooling curve, before climbing back up to the compression luminosity once mass transfer begins again.
During these same periods, after a burst but before re-contact, the donor luminosity increases.
This happens because during mass loss periods, the nuclear burning luminosity in the donors core is absorbed in the stellar envelope (\cite{Yungelson2008}).
In the donor's envelope there is no nuclear burning, so the luminosity's slope is dominated by mass loss ($dL/dm\approx(\dot{M}/\Delta M)(k_BT/m_p)<0$, where $\Delta M$ is overlying mass).
When mass loss suddenly shuts off, this mass loss term disappears giving $dL/dm\approx0$ in the envelope, meaning the trapped heat gets released on the timescale $\sim10^4$ years.

The features seen in the bottom two panels after $16$ Myr result from the exposure of the once convective helium burning core.
When sharp compositional gradients are exposed, the mass transfer rates experience a spike.

For all cases studied here, except the one with $M_{\rm WD}=1.26 M_\odot$, the first burst is the largest because the helium is accumulated at the lowest rate.
As the mass transfer rates decrease after the period minimum, they quickly become too low to trigger helium ignitions (\cite{Shen2009}) and the WD ends up with a cold, thick helium shell.
The case with the $M_{\rm WD}=1.26 M_\odot$ accretor is the only case in which the last explosion is the strongest (Figure \ref{fig:8}), as the ignition mass is much lower for the more massive WD accretors. 
In fact, the last explosion on the $M_{\rm WD}=1.26 M_\odot$ accretor is the most powerful one studied here.
After this last explosion, likely as a .Ia SN, the donor does not have enough mass left to increase the mass of the accretor to the Chandrasekhar mass, so an accretion induced collapse is not expected through this channel.

\section{Conclusions}\label{sec:conc}

We present the first self-consistent calculation of mass transfer between a low-mass helium star and a WD in a short-period binary and the resulting helium shells and their explosions on the surface of the WD.
We started by simulating the secular evolution of low mass helium stars, that is, evolved with a point-mass binary companion, to see how structure and mass-loss feed back on each other.
We found, confirming \cite{Yungelson2008}, that during the first stage of mass transfer, the donor maintains thermal equilibrium, as $\tau_{\rm KH}<\tau_{\dot{M}}$, leading to a relatively constant $\dot{M}$ during the so-called plateau phase.
Eventually the donor loses enough mass to effectively shut-off nuclear burning and reverse the timescale inequality, meaning the star responds adiabatically to mass-loss.
At this point mass transfer rates rise until mass transfer effects become dominant over GWR in determining the orbital separation, and the system reaches a period minimum of 9-10 minutes, after which the binary separation increases.
Core evolution proceeds adiabatically along nearly constant entropy trajectories that are mildly degenerate until nearly all the mass of the donor has been removed.

When we include the evolution of the accreting WD in the binary simulations, helium shells build up on the WD until a thermonuclear runaway develops.
In studying these models, we find that the first flash is usually the largest, and the first flashes from systems with $M_{\rm He}>0.4 M_\odot$ and $M_{\rm WD}>1.0 M_\odot$ meet the requirements for triggering dynamical burning and developing a deflagration or detonation.
Using the conditions derived by \cite{Moore2013} and \cite{Shen2014}, all the flashes studied here, including the ``later'' flashes, are massive enough to sustain a laterally propagating detonation within the He shell.
Furthermore, the first bursts from the $0.8 M_\odot$ and the $1.0 M_\odot$ WD accretors, and all the burst from the $1.26 M_\odot$ WD accretor, will produce a significant fraction of isotopes that are radioactive on relevant timescales. 

We also follow through with some simulations assuming that the first helium flash does not unbind the WD.
The ejected mass increases the binary separation, temporarily shutting off mass transfer until GWR brings the stars back into contact.
This series of events repeats itself until the stars are far enough apart and mass transfer rates are low enough to prevent runaways in the helium shells.

We thank Kevin Moore for helpful discusions regarding helium detonations, along with Dean Townsley and Ken Shen for their useful comments.
We would also like to thank the referee for their many helpful comments.  
This work was supported by the National Science Foundation under grants PHY 11-25915, AST 11-09174, and AST 12-05574. 
Most of the simulations for this work were made possible by the Triton Resource, a high-performance research computing system operated by the San Diego Supercomputer Center at UC San Diego.

\bibliographystyle{apj}
\bibliography{lowmass_donors_bib}

\end{document}